\documentstyle[aps,preprint,epsf]{revtex}
\begin{document}

\newcommand{\half}   {\mbox{${\textstyle \frac{1}{2}}$}}
\newcommand{\beq}    {\begin{equation}}
\newcommand{\eeq}    {\end{equation}}
\newcommand{\beqa}   {\begin{eqnarray}}
\newcommand{\eeqa}   {\end{eqnarray}}

\title{On the semiclassical theory for universal transmission \\
       fluctuations in chaotic systems: the importance of unitarity} 

\author{Ra\'ul O. Vallejos and Caio H. Lewenkopf}

\address{Instituto de F\'{\i}sica,
	 Universidade do Estado do Rio de Janeiro \\
	 R. S\~ao Francisco Xavier, 524, 
       CEP 20559-900 Rio de Janeiro, Brazil}

\date{\today}

\maketitle

\begin{abstract}
The standard semiclassical calculation of transmission correlation 
functions for chaotic systems is severely influenced by unitarity problems.
We show that unitarity alone imposes a set of relationships between 
cross sections correlation functions which go beyond the 
diagonal approximation.
When these relationships are properly used to supplement the 
semiclassical scheme we obtain transmission correlation functions
 in full agreement 
with the exact statistical theory and the experiment. 
Our approach also provides a novel prediction for the transmission 
correlations in the case where time reversal symmetry is present. 
\end{abstract}

\draft\pacs{}

\narrowtext

\section{Introduction}

Since the pioneering work of Bl\"umel and Smilansky \cite{Blumel88}
the semiclassical $S$-matrix \cite{Miller} has been used by many authors 
to study fundamental questions related to quantum chaotic scattering. 
In recent years the interest in this subject has grown due to 
the experimental investigation of electronic transport through 
small devices, such as open quantum dots \cite{Marcus}.
At sufficiently low temperatures, these devices preserve quantum
coherence and are called mesoscopic.
The conductance in transport processes that preserve quantum coherence 
is directly related to the $S$-matrix by the Landauer-B\"uttiker formula 
\cite{Datta95}.
Since the classical underlying electronic dynamics in quantum
dots is believed to be chaotic, these are excellent systems
to observe the quantum manifestations of classical chaotic 
scattering \cite{Baranger96}.

One of the central issues in mesoscopic physics is to single out
statistical universal properties of quantum systems.
In this paper we show that this goal cannot be theoretically 
achieved using the standard semiclassical 
$S$-matrix approach, since the latter is not  
able to provide trustworthy results for universal 
cross-section or conductance correlation functions. We also show
that this situation can be fixed by imposing a set of
semiclassical sum rules that guarantee unitarity.
 
For a device connected to reservoirs by two leads, the 
Landauer-B\"uttiker formula relates its conductance $G$
to the transmission coefficient $T$ by the expression
$G = (2e^2/\hbar)T$. In electronic transport, $T$ is usually called the
dimensionless conductance.
When the entrance and exit leads support $N_1$ and $N_2$ propagating 
modes or channels, respectively,
the transmission $T$ reads
\begin{equation}
T(x) = \sum_{a=1      }^{N_1} \; 
       \sum_{b=N_1 + 1}^N 
       |S_{ab}(x)|^2 
     = \sum_{a,b} \sigma_{ab}(x) \; ,
\end{equation}
where we have introduced $\sigma_{ab}$, the transition probability 
(the cross-section, apart from a kinematical factor) from channel $b$ in one lead to channel 
$a$ in the other lead. 
The parameter $x$ represents either the energy $E$, or an external 
parameter $X$ such as a magnetic field, or both.  
Without loss of generality we restrict our discussion to the 
case where $N_1= N_2  = N/2 \equiv M$.

The original problem of conductance fluctuations in a 
mesoscopic device is now cast into a more generic one of 
transmission fluctuations of a quantum process where a particle is
chaotically scattered.
Thus, this discussion is also of interest for transmission experiments
with irregular microwave cavities, ``chaotic" atoms and nuclei. 
Our goal is to use the semiclassical theory to describe 
the statistical properties of the transmission as the parameter 
$x$ is varied. 
This information is contained in the average value of $T$ (a 
two-point statistical measure of the $S$-matrix elements) and
in its autocorrelation function (a four-point function), defined 
as
\begin{equation}
(T,T^\prime) \equiv 
       \langle T(x) T(x^\prime) \rangle -  
       \langle T(x) \rangle \langle T(x^\prime) \rangle
  \; .
\end{equation}
The average is taken over $x$ and $x'$ keeping the difference 
$|x-x'|$ fixed.
The transmission autocorrelation function is directly related
to the covariances of the transition probabilities
\begin{equation}
\label{eq:TT'}
(T,T^\prime) = \sum_{ a,c=  1}^M 
               \sum_{~b,d=M+1}^N 
	      (\sigma_{ab},\sigma^\prime_{cd}) \; .
\end{equation}

The variance of $T$, var($T$), is the statistical measure of a 
fundamental phenomenon in mesoscopic physics: 
For systems where quantum coherence is preserved the dimensionless 
conductance displays fluctuations of order unity irrespective of 
sample size, provided the dynamics is chaotic (or diffusive) and 
there are no tunneling barriers hindering the transmission. 
This phenomenon is known as ``universal conductance fluctuations'' 
(UCF) \cite{Altshuler}.
Thus, a successful approximation scheme to explain UCF has to be 
accurate to the level of unity for the variance of $T$.
In the specific case of quantum dots, {\sl i.e.}, ballistic electronic 
cavities, the random matrix theory \cite{Mello94,Pichard94} and the
supersymmetric method \cite{Efetov95} are, so far, 
the successful approaches to calculate var($T$).

The purpose of this work is twofold. 
First we analyze a very simple statistical measure -- the average 
cross section -- to show that the standard semiclassical $S$-matrix 
theory does not achieve the required precision to be useful as a 
theory for UCF. 
In doing this, we indicate its main sources of inaccuracy and discuss 
the main problems involved in improving the theory.
We then show how to fix the inaccuracies by making explicit use of 
the unitarity of the $S$ matrix.
This procedure is similar in spirit to those used in semiclassical 
studies of spectra of closed system \cite{Bogomolny92}. 
It can be viewed as a proposal for a set of semiclassical scattering 
sum rules to enforce well known exact symmetries of the $S$-matrix.

\section{The semiclassical approach}
\label{sectionII}

We start with Miller's semiclassical $S$-matrix formula \cite{Miller}
 (including now transmission and reflections)
\beq
\label{eq:Miller}
    \widetilde{S}_{ab}(E, X) = 
    \sum_{\mu(a,b)} \sqrt{p_\mu(E, X)} \, 
    e^{i\phi_\mu(E, X)/\hbar} \;,
\eeq
where $\mu(a, b)$ labels the classical trajectories that 
start at channel $b$ and end at channel $a$, $\phi_\mu$
are their reduced actions (with a Maslov phase included), and 
$p_\mu$ stands for the classical transition probability of
going from $a$ to $b$ through the orbit $\mu$ \cite{Smilansky}
 (here and throughout the paper the tilde indicates that the
semiclassical approximation is employed). 
It is implicit in the derivation of Eq.~(\ref{eq:Miller}) that the 
number of open channels must be much larger than one and that there 
are no tunneling barriers between the leads and the cavity. 
When the scattering is chaotic (and the short time dynamics does
not significantly contribute to $S$) the domains of applicability 
of the semiclassical theory and random matrix theory coincide and 
both approaches should be comparable. 
In this regime the semiclassical approach provides the dynamical 
explanation for the universality of the scattering fluctuations. 

In general the semiclassical $S$-matrix is not exactly unitary at any 
given energy $E$. 
Indeed, it is only upon energy averaging and for the 
case of broken time-reversal symmetry (BTRS), that unitarity 
is automatically fulfilled.
We now show this known result so as to present the basic elements
and approximations employed in this paper.
The energy averaged semiclassical cross section reads
\beq
     \langle \widetilde{\sigma}_{ab} \rangle  
  =  \langle |\widetilde{S}_{ab}|^2 \rangle 
  =  \sum_{\mu,\nu} 
     \left\langle 
     \sqrt{p_\mu p_\nu} \, 
      e^{i(\phi_\mu - \phi_\nu)/\hbar} 
     \right\rangle \; .
\eeq
Here $\langle \cdots \rangle$ indicates an energy average
within an energy window where the classical dynamics presents little
changes, nonetheless comprising many resonances. 
To compute the energy average one neglects the energy dependence of
the probabilities $p_\mu$ and uses the diagonal approximation.
The latter says that, on average, only orbits having the same action are
correlated. If there are no symmetries present, this means that 
$ \left\langle \exp[i(\phi_\mu - \phi_\nu)/\hbar] 
  \right\rangle = \delta_{\mu \nu}$. 
Then
\beq
 \langle \widetilde{\sigma}_{ab} \rangle =
 \sum_{\mu,\nu} \sqrt{p_\mu p_\nu} \, \delta_{\mu \nu}
 = \sum_{\mu} p_\mu                        \; .
\eeq
The proof is completed by using the classical normalization condition
\cite{Smilansky} 
\beq
\sum_{a=1}^N \sum_{\mu(a,b)} p_\mu = 1 \;,
\eeq
which insures that 
$\sum_a \langle |\widetilde{S}_{ab}|^2 \rangle = 1$.

We shall assume that for any given entrance channel $b$, all exit 
channels $a$ are equivalent, {\sl i.e.},
\beq
\sum_{\mu(a,b)} p_\mu = \frac{1}{N}\; ,
\eeq
which yields 
\beq
 \langle \widetilde{\sigma}_{ab} \rangle =  \frac{1}{N} \; .
\eeq
The assumption of equivalent channels is justified (in the BTRS case)
if the particle typically stays inside the interaction region time 
enough to be randomized, meaning that it becomes equiprobable
to be ejected through any outgoing channel. 
The analysis of the other limiting case where time reversal symmetry is 
preserved will be postponed to Section \ref{sectionIII}.

In order to clearly explain why unitarity problems affect the 
semiclassical theory of transmission fluctuations we introduce the 
object
\beqa
\widetilde{1}_a \equiv 
\sum_{b=1}^N \widetilde{\sigma}_{ab} = 
\sum_{b=1}^N \sum_{{\mu(a,b)}\atop{\nu(a,b)}}
     \sqrt{p_\mu p_\nu} \, e^{i(\phi_\mu - \phi_\nu)/\hbar} \; .
\eeqa
If the semiclassical $S$-matrix had been exact $\widetilde{1}_a=1$;
instead one has $\langle \widetilde{1}_a \rangle =1$.
The lack of precision of the standard semiclassical scattering 
theory at the four-point level becomes evident by analyzing
the variance $(\widetilde{1}_a, \widetilde{1}_b)$.
The semiclassical approximation gives 
$(\widetilde{1}_a, \widetilde{1}_b) \neq 0$, 
leading to a ``unit fluctuation" problem. To see this, 
using Eq.~(\ref{eq:Miller}),  we write
\beq
\left( \widetilde{1}_a, \widetilde{1}_b \right) 
= \left\langle \sum_{c,d=1}^N
  \sum_{{\mu(a,c)}\atop{\nu(a,c)}}
  \sum_{{\mu^\prime(b,d)}\atop{\nu^\prime(b,d)}}
  \sqrt{p_\mu p_\nu p_{\mu^\prime} p_{\nu^\prime}}\,
  \exp\Big[ \frac{i}{\hbar}(\phi_\mu          - \phi_\nu +
                            \phi_{\mu^\prime} - \phi_{\nu^\prime})
  \Big] \right\rangle - 1
\eeq
so that the diagonal approximation yields
\beq
\label{eq:1sqinter}
\left( \widetilde{1}_a, \widetilde{1}_b  \right) =  \sum_{c,d=1}^N
  \sum_{{\mu(a,c)}\atop{\nu(a,c)}}
  \sum_{{\mu^\prime(b,d)}\atop{\nu^\prime(b,d)}}
  \sqrt{p_\mu p_\nu p_{\mu^\prime} p_{\nu^\prime}}
  (\delta_{\mu\nu}\delta_{\nu^\prime\mu^\prime} +
   \delta_{\mu\nu^\prime}\delta_{\nu\mu^\prime} ) -1\;.
\eeq
The first Kronecker $\delta$ product decouples the
sums over orbits starting at channel $c$ and ending at $a$ from 
those entering the scattering region through channel $d$ and
exiting through $b$. 
The resulting double sum adds up to unity.
The second product contains crossed terms which vanish unless
$a=b$ and $c=d$. Equation (\ref{eq:1sqinter}) becomes
\beq
\label{eq:flucofone}
\left( \widetilde{1}_a, \widetilde{1}_b  \right) 
  = \delta_{ab} \sum_{c=1}^N \sum_{{\mu(a,c)}\atop{\nu(a,c)}}
    p_\mu p_\nu 
  = \frac{\delta_{ab}}{N} \;.
\eeq
This inaccuracy is neither unexpected, nor large. 
However, it has important consequences 
for the calculation of transmission fluctuations.
This becomes evident by inspecting
\beq
\left( \sum_{a=1}^N \widetilde{1}_{a}, 
       \sum_{b=1}^N \widetilde{1}_{b}
\right) =
\left( \sum_{a,c=1}^N \widetilde{\sigma}_{ac}, 
       \sum_{b,d=1}^N \widetilde{\sigma}_{bd} 
\right) = 1 \neq 0
\eeq
which has the same double sum structure of the transmission 
variance and shows an inaccuracy exactly of the order
of the effect that we aim to describe.

Let us be more explicit and go back to the analysis of 
the transmission autocorrelation function. 
Recalling Eq.~(\ref{eq:TT'}) and assuming channels 
to be statistically equivalent we write
\begin{equation}
\label{eq:TT'explicit}
(T,T^\prime) =   M^2         (\sigma_{ab},\sigma^\prime_{ab}) +
               2 M^2 (M-1)   (\sigma_{ab},\sigma^\prime_{ac}) +
	         M^2 (M-1)^2 (\sigma_{ab},\sigma^\prime_{cd})     \; .
\end{equation}
Here we use the convention that different indices $a$ and $b$ in the 
covariances imply that $a \neq b$.
This means that for the above equation $b \ne c$ in the second 
term of its RHS, and $a \ne c$ and $b \ne d$ for the third one.
We shall demonstrate below that, owing to unitarity, all three 
terms in the RHS of Eq.~(\ref{eq:TT'explicit}) are of the same 
order of magnitude.
However, within the diagonal approximation,
both the semiclassical covariances
$(\widetilde{\sigma}_{ab},\widetilde{\sigma}'_{ac})$ and 
$(\widetilde{\sigma}_{ab},\widetilde{\sigma}'_{cd})$ are zero. 
Let us admit that the semiclassical approach gives the correct
result for the transmission probability covariances to order
$1/N^2$. 
Then, for a successful description of the transmission 
fluctuations, the theory has to be improved to access 
the first non vanishing order in the non diagonal terms 
$(\sigma_{ab},\sigma'_{ac})$ and $(\sigma_{ab},\sigma'_{cd})$,
which are ${\cal O}(N^{-3})$ and ${\cal O}(N^{-4})$, respectively.
Though desirable this is not really necessary.
The alternative scheme we propose is to bypass the explicit semiclassical 
calculation of the nondiagonal covariances and use the unitarity of the 
$S$ matrix to relate 
the latter covariances to the diagonal one. Having expressed 
$(T,T')$ in terms of $(\sigma_{ab},\sigma'_{ab})$ alone, we use the 
semiclassical approximation only at the very end.

\section{Enforcing unitarity}
\label{sectionIII}

The relations among diagonal and nondiagonal covariances can be easily 
obtained from 
\begin{equation}
(\sum_{b=1}^N \sigma_{ab},\sigma'_{cd}) = (1, \sigma'_{cd}) =0   \; ,
\end{equation}
which follows from unitarity. 
To proceed further we have to separately analyze the cases where
either time reversal symmetry is absent (BTRS) or present (TRS).
This distinction is necessary because the ``elastic'' processes ($a=b$) 
and the ``inelastic'' ones ($a \neq b$) display different statistical 
properties when time reversal symmetry is preserved. 
Indeed it is well 
known that due to quantum interference, time reversal symmetry enhances
the average reflection 
probability $\langle \sigma_{aa} \rangle$ by a factor of 
two \cite{Mahaux79}. This can be understood semiclassically by noting that
in the TRS case there are pairs of orbits having the same action
(time reversal partners) which contribute to $\sigma_{aa}$, 
thus interfering constructively to produce the factor two.
Due to this effect the classical equivalence of channels breaks down 
at the quantum level. 
In this case the equivalence is only restored when time reversal symmetry
is broken, leading to the considerations presented in Section \ref{sectionII}.

To illustrate the consequences of this phenomenon
and the spirit of our scheme, let us analyze $\langle T(x) \rangle$ in
the crossover regime from TRS to BTRS.
Recall that 
\begin{equation}
\langle T(x) \rangle = \sum_{a=1}^M \sum_{~b=M+1}^N 
                       \langle \sigma_{ab}(x) \rangle
                     = M^2       \langle \sigma_{ab}(x) \rangle .
\end{equation}
Here $x$ parameterizes a Hamiltonian change breaking  
time reversal symmetry as $x$ grows from zero (TRS) to some critical 
value $x^\ast$ (BTRS).
Unitarity relates diagonal and off-diagonal averages:
\begin{equation}
\label{unitsigma}
1 =       \langle \sigma_{aa}(x) \rangle + 
    (N-1) \langle \sigma_{ab}(x) \rangle    \; .
\end{equation}
This equation allows us to write $\langle T \rangle$ in terms 
of $\langle \sigma_{aa} \rangle$, the average that is semiclassically
sensitive to time reversal effects, to obtain 
\begin{equation}
\label{Tweak}
\langle T(x) \rangle = \frac{N^2}{4(N-1)} 
                       (1-\langle \sigma_{aa}(x) \rangle)  \; .
\end{equation}
For $x=0$ the elastic enhancement is maximal and hence 
$\langle T(x) \rangle$ takes its smallest value. In mesoscopic
physics, to distinguish from strong localization which is a 
phenomenon very different in origin, the reduction of transmission 
due to TRS is called the weak localization peak. 
Up to this point Eq.~(\ref{Tweak}) is an exact expression. 
The semiclassical result is obtained by calculating 
$\langle \sigma_{aa}(x) \rangle$ from Miller's formula. 
For $x$ representing 
a magnetic field, the semiclassical approach gives a Lorentzian 
shape for the weak localization peak \cite{Baranger93a} in agreement
with RMT \cite{Weidenmuller94}. 
The amplitude of the weak localization correction can be readily 
obtained recalling that
$\langle \widetilde{\sigma}_{aa}(0) \rangle \approx 2/N$ 
and 
$\langle \widetilde{\sigma}_{aa}(x \ge x^\ast) \rangle \approx 1/N$, 
so that
\begin{equation}
  \langle \widetilde{T}(0)       \rangle - 
  \langle \widetilde{T}(x^\ast) \rangle   = - \frac{1}{4} \; ,
\end{equation}
again in agreement with random matrix theory \cite{Mello94}.
The discussion above is not entirely original and was inspired 
by the pioneer semiclassical study of the weak localization peak 
in ballistic cavities developed by Baranger and collaborators
\cite{Baranger93a}.
Based on the same strategy presented above, we are now ready
to understand the UCF problem.

\subsection{Transmission fluctuations in systems with broken time 
            reversal symmetry}

In this case all $S$ matrix elements are statistically equivalent. 
In order to express $(T,T')$ in terms of $(\sigma_{ab},\sigma'_{ab})$
it suffices to consider the following two independent unitarity 
equations
\beq
\label{unit0}
\sum_{b=1}^N (\sigma_{ab},\sigma'_{ac}) = 0 \; ,~~~ 
\sum_{b=1}^N (\sigma_{ab},\sigma'_{cd}) = 0 \; 
\eeq
($c \ne a$ in the second equation).
The above relations can be reduced to:
\beq
\label{unit}
            (\sigma_{ab},\sigma'_{ab}) + 
      (N-1) (\sigma_{ab},\sigma'_{ac})    
                =  0  \; ,~~~  
            (\sigma_{ab},\sigma'_{ac}) + 
      (N-1) (\sigma_{ab},\sigma'_{cd})    
                =  0  \;   
\eeq
(the convention about indices being the same as in 
Eq.~(\ref{eq:TT'explicit})).
Notice that these relations are not satisfied in the diagonal
approximation. At the semiclassical level Eqs.~(\ref{unit0}) and 
(\ref{unit}) can be regarded as sum rules that go beyond the diagonal 
approximation. 
Insertion of Eqs.~(\ref{unit}) into Eq.~(\ref{eq:TT'explicit}) 
readily renders
\begin{equation}
\label{ttGUE}
(T,T')= \frac{M^4}{(2M-1)^2} (\sigma_{ab},\sigma'_{ab}) \; .
\end{equation}
As in the two-point analysis this equation is exact.

Now we are ready to employ the semiclassical approximation to compute
$(\sigma_{ab},\sigma'_{ab})$.
Let us first consider the case where $x$ stands for the 
energy, {\em i.e.}, 
\beq
\label{Cab}
(\sigma_{ab},\sigma'_{ab}) = C_{ab}(\varepsilon) \equiv 
    \left  \langle  \sigma_{ab}(E+\frac{\varepsilon}{2})
                 \; \sigma_{ab}(E-\frac{\varepsilon}{2}) 
    \right \rangle_E - 
    \left  \langle  \sigma_{ab}
    \right \rangle_E^2 \; .
\eeq
The semiclassical autocorrelation function $\widetilde{C}_{ab}(\varepsilon)$ 
can be calculated for classically small values of $\varepsilon$, 
{\sl i.e.}, for energy differences such that the classical perturbation 
theory holds.
In this case one keeps the stability coefficients
constant and expands the actions to first order
in $\varepsilon$, {\em i.e.},
$\phi_\mu(E  \pm \varepsilon/2) \approx 
 \phi_\mu(E) \pm \tau_\mu \varepsilon/2$; here
$\tau_\mu$ is the time the particle takes to travel from channel $b$
to channel $a$ along the orbit $\mu$.
After the diagonal approximation we obtain
\beqa
\left \langle   \widetilde{\sigma}_{ab}(E+\frac{\varepsilon}{2})
             \; \widetilde{\sigma}_{ab}(E-\frac{\varepsilon}{2})
\right\rangle =
  \sum_{{\mu       (a,b)} \atop{\nu       (a,b)}} 
  \sum_{{\mu^\prime(a,b)} \atop{\nu^\prime(a,b)}}  &&
  \sqrt{p_\mu p_\nu p_{\mu^\prime} p_{\nu^\prime}}
  (\delta_{\mu\nu}        \delta_{\nu^\prime\mu^\prime} +
   \delta_{\mu\nu^\prime} \delta_{\nu\mu^\prime} 
   ) \nonumber\\                                  && 
  \times\exp \left[ 
  \frac{i \varepsilon}{2 \hbar}
  \left( \tau_\mu - \tau_\nu + 
         \tau_{\mu^\prime} - \tau_{\nu^\prime} \right)
             \right]  \;.
\eeqa
Using the same arguments employed after 
Eq.~(\ref{eq:1sqinter}) we arrive at
\beq
\widetilde{C}_{ab}(\varepsilon) =
  \sum_{{\mu(a,b)}\atop{\nu(a,b)}} p_\mu p_\nu
  \exp \left[ i\frac{\varepsilon}{\hbar}(\tau_\mu - \tau_\nu)\right]
   \;.
\eeq
According to the analogue of the Hannay-Ozorio de Almeida sum rule 
for open systems \cite{Kadanoff84},  
\beq
   \sum_{t\le \tau_\mu \le t+\delta t} p_\mu  = 
   \frac{\gamma}{N}e^{-\gamma t} 
   \delta t\;,
\eeq
where 
$\sum_{t\le \tau_\mu \le t+\delta t} p_\mu$ 
is the sum of all classical transition probabilities from channel 
$a$ to $b$ through trajectories within a small time interval 
$[t, t+\delta t]$, where $\delta t$ is classically small. 
The exponential is determined by the inverse escape time $\gamma$, 
later to be associated to an energy width $\Gamma = \hbar \gamma$.
Replacing the sum over orbits by an integral over the time, we 
finally obtain 
\beq
\label{eq:corE}
\widetilde{C}_{ab}(\varepsilon) = 
 \frac{1}{N^2} \, \frac{1}{1 + (\varepsilon/\Gamma)^2} \;.
\eeq

Before commenting on this result, we shall generalize it 
by accounting for an external parametric change $X$ in the 
Hamiltonian.
For instance, in (mesoscopic physics) experiments $X$ is frequently 
an external magnetic field.
The strategy to compute the generalized correlation function 
is, as above, based on classical perturbation theory. By expanding
the reduced action to first order in $X$, we have to deal with
$Q_\mu \equiv \partial \phi_\mu/\partial X$.
A full account of the technical details involved in calculating
the parametric correlation can be found in Ref.~\onlinecite{Ozorio98}.
The basic step though is to compute the time average
\beq
 \left  \langle e^{iQ_\mu \delta X/\hbar} 
 \right \rangle_{\delta t} =
 \exp   \left[ - \frac{\delta X^2}{2\hbar^2} 
                \langle Q^2(t) \rangle_{\delta t}
        \right] \;.
\eeq
Since $\langle Q^2(t) \rangle_{\delta t}$ grows diffusively with time, 
{\sl i.e.}, $\langle Q^2(t) \rangle_{\delta t} = \alpha t$, the 
autocorrelation function becomes
\beq
 \widetilde{C}_{ab}(\varepsilon, \delta X) = 
 \frac{1}{N^2} 
 \frac{1}{[ 1 + (\delta X/X_c)^2 ]^2 + (\varepsilon/\Gamma)^2}\;,
\eeq
with 
\beq
\label{eqalfa}
X_c^2 \equiv 2\hbar\Gamma/\alpha \; .
\eeq

Then the semiclassical result that includes unitarity restrictions 
is 
\begin{equation}
 (\widetilde{T},\widetilde{T}^\prime) = 
 \frac{1}{16} 
 \frac{1}{[ 1 + (\delta X/X_c)^2 ]^2 + (\varepsilon/\Gamma)^2} + {\cal O} (1/N) \; .
\eeq
Remarkably, this result agrees exactly with the dimensionless 
conductance autocorrelation function for open ballistic dots in 
the limit $N \gg 1$ obtained by Efetov \cite{Efetov95} using the 
supersymmetric technique.
The agreement extends also to the structure of the parameter
$X_c$ (\ref{eqalfa}) if one relates $\alpha$ to the level velocity
of closed systems as defined in Ref.~\cite{Ozorio98}.
In this respect the semiclassical approach is complementary to
random matrix theories, since it provides a dynamical interpretation
for the nonuniversal quantities $X_c$ and $\Gamma$.

\subsection{Systems with time reversal symmetry}

Now the $S$ matrix is symmetric and the statistical
properties of diagonal and off-diagonal elements are different.
Hence we need to write down two additional unitarity relations, as 
compared with the BTRS case, in order to single out the elastic case 
separately:
\beq
\sum_{b=1}^N (\sigma_{ab},\sigma'_{aa}) = 0 \; , ~~
\sum_{b=1}^N (\sigma_{cb},\sigma'_{aa}) = 0 \; , ~~ 
\sum_{b=1}^N (\sigma_{ab},\sigma'_{ac}) = 0 \; , ~~ 
\sum_{b=1}^N (\sigma_{ab},\sigma'_{cd}) = 0 \; 
\eeq
($c,d \ne a$). 
In terms of the basic covariances, the above system of equations is 
rewritten as
\begin{eqnarray}
            (\sigma_{aa},\sigma'_{aa}) + 
(N-1) (\sigma_{aa},\sigma'_{ab})        & = & 0   \nonumber\\
            (\sigma_{aa},\sigma'_{ab}) + 
            (\sigma_{aa},\sigma'_{bb}) + 
(N-2) (\sigma_{aa},\sigma'_{bc})        & = & 0   \nonumber\\
            (\sigma_{ab},\sigma'_{ab}) + 
            (\sigma_{aa},\sigma'_{ab}) + 
(N-2) (\sigma_{ab},\sigma'_{ac})        & = & 0   \nonumber\\
            (\sigma_{aa},\sigma'_{bc}) + 
          2  (\sigma_{ab},\sigma'_{ac}) + 
(N-3) (\sigma_{ab},\sigma'_{cd})        & = & 0    \; ,
\label{eqstrs}
\end{eqnarray}
where the channel index convention is that following 
Eq.~(\ref{eq:TT'explicit}). 
The only difference with the BTRS case is the factor 2 in the last
equation, which is a consequence of the symmetry of $S$.
As in the preceding subsection, we would like to express all 
covariances in 
terms of
$(\sigma_{aa},\sigma'_{aa})$ and
$(\sigma_{ab},\sigma'_{ab})$ which are the only nonzero ones 
in the diagonal approximation. Regrettably, we have 
more unknowns than equations, and it is not possible to obtain an
exact equation like Eq.~(\ref{ttGUE}). However, as we are only interested in
a relation which is correct to leading order in $1/N$, it suffices
to consider the simplified system
\beq
        (\sigma_{ab},\sigma'_{ab}) + 
      N (\sigma_{ab},\sigma'_{ac})    =  {\cal O}(N^{-3})  \; ,~~~
      2 (\sigma_{ab},\sigma'_{ac}) + 
      N (\sigma_{ab},\sigma'_{cd})    =  {\cal O}(N^{-4})   \; ,
\eeq
which is obtained from the last two equations in (\ref{eqstrs}) by 
keeping only the leading terms. 
These relations are the TRS analogues of (\ref{unit}) and 
lead to 
\begin{equation}
(T,T^\prime) \approx \frac{M^2}{2} (\sigma_{ab},\sigma'_{ab}) \; .
\end{equation}
Semiclassically, time reversal effects only manifest themselves in the 
diagonal covariances $(\sigma_{aa},\sigma'_{aa})$. The correlator 
$(\widetilde{\sigma}_{ab},\widetilde{\sigma}'_{ab})$ is the same as that 
for the BTRS case. Thus, at the semiclassical level,
the effect of time reversal symmetry is to enhance the transmission 
fluctuations by a factor of two (cf. Eq.~(\ref{ttGUE})), without changing 
the shape of the correlation function, that is
\begin{equation}
 (\widetilde{T},\widetilde{T}^\prime) =
 \frac{1}{8} 
 \frac{1}{[ 1 + (\delta X/X_c)^2 ]^2 + (\varepsilon/\Gamma)^2} + {\cal O} (1/N) \; .
\eeq
We are not aware of any study of the transmission autocorrelation function 
for the TRS case using the supersymmetric method. 
For technical reasons it is much simpler to obtain the variance var$(T)$ 
using random matrix theory. This result is known \cite{Mello94} and, 
for $N \gg 1$, agrees with our semiclassical calculation
\begin{equation}
\mbox{var}(\widetilde{T}) = \frac{1}{8} \; .
\end{equation}

\section{Concluding Remarks}

Spectral studies in closed system have shown
that the diagonal approximation should start to fail when the orbits 
involved have periods of the order of the Heisenberg time $\tau_H$. 
In scattering systems, the contribution of orbits with periods 
larger than the mean escape time $\tau_e$ is negligible.
Given that in the semiclassical regime $\tau_e \ll \tau_H$ 
it is generally accepted that the diagonal approximation should 
be unproblematic for scattering systems. 

However, we have shown that the standard semiclassical approach fails to 
describe
the transmission fluctuations because the diagonal approximation does not 
preserve the unitarity of the $S$ matrix to the required precision.
One way to circumvent this problem, perhaps the most satisfactory from 
a theoretical point of view, is to improve the semiclassical theory to 
include correlations between different orbits. 
Alternatively, we have shown that the unitarity of the $S$ matrix can be 
used to express the transmission autocorrelation function in terms of
transmission probabilities.
Such expression contains some information about unitarity allowing the 
standard semiclassical approximation to be invoked resulting in a theory
consistent with UCF.

Other difficulties are also encountered when calculating the
transmission correlations: The semiclassical transmission 
correlator is not translationally invariant.  
For instance, this is manifest in the fact that 
\beq
 \langle \widetilde{T}(E) 
         \widetilde{T}(E + \varepsilon)   \rangle \neq 
 \langle \widetilde{T}(E + \varepsilon/2) 
         \widetilde{T}(E - \varepsilon/2) \rangle  \; ,
\eeq
which can easily be checked by inspection.
In our calculations we preferred to use 
$\langle \widetilde{T}(E + \varepsilon/2) 
         \widetilde{T}(E - \varepsilon/2) \rangle $
because it is explicitly real. This choice is consistent with the spirit 
of this work, {\em i.e.}, all information about exact quantum symmetries must
be used in trying to compensate for the shortcomings of the semiclassical
$S$ matrix.

\acknowledgements

The authors gratefully acknowledge discussions with A. M. Ozorio de 
Almeida and O. Bohigas. This work was partially supported by CNPq and 
PRONEX (Brazil).


\begin{thebibliography}{99}

\bibitem{Blumel88}
  Bl\"umel R and Smilansky U 1988
     {\it Phys. Rev. Lett.} {\bf 60} 477

\bibitem{Miller}
  Miller W H 1975
     {\it Advances in Chemical Physics} ed K. P. Lawley
     (New York: Wiley) Vol. 30 p 77.

\bibitem{Marcus}
  Marcus C M et. al. 1992
     {\it Phys. Rev. Lett.} {\bf 69}, 506 
  
\bibitem{Datta95}                 
  Datta S 1995 
    {\it Electronic Transport in Mesoscopic Physics} (Cambridge University Press)
  
\bibitem{Baranger96}
  Jalabert R A, Baranger H U and Stone A D 1990
     {\it Phys. Rev. Lett.} {\bf 65} 2442

\bibitem{Altshuler}
  Altshuler B L and Shklovskii B I 1986
     {\it Sov. Phys. JETP} {\bf 64} 127 

\bibitem{Mello94}
  Baranger H U and Mello P A 1994
     {\it Phys. Rev. Lett.} {\bf 73} 142
     
\bibitem{Pichard94}
  Jalabert R A, Pichard J-L and Beenakker C W J 1994
     {\it Europhys. Lett.} {\bf 27} 255 

\bibitem{Efetov95}
  Efetov K B 1995
     {\it Phys. Rev. Lett.} {\bf 74} 2299     

\bibitem{Bogomolny92}
  Bogomolny E 1992
     {\it Nonlinerity} {\bf 5 } 805  \\
  Smilansky U 2000
     {\it J. Phys. A: Math. Gen.} {\bf 33} 2299
     
\bibitem{Smilansky}
  Smilansky U 1991 {\it Les Houches 1989 Session LII on Chaos and 
          Quantum Physics} ed  M J Gianonni, A Voros and J Zinn-Justin 
     (Amsterdam: North-Holland) pp 371-441     
     
\bibitem{Mahaux79}
  Mahaux C and Weidenm\"uller H A 1979 
      {\it Ann. Rev. Nucl. Sci.} {\bf 29} 1

\bibitem{Baranger93a}
  Baranger H U, Jalabert R A and Stone A D 1993
     {\it Phys. Rev. Lett.} {\bf 70} 3876 \\
  Baranger H U, Jalabert R A and Stone A D 1993
     {\it CHAOS} {\bf 3} 665
     
\bibitem{Weidenmuller94}
  Pluha\v{r} Z, Weidenm\"uller H A, Zuk J A, Lewenkopf C H and Wegner F J 1995
     {\it Ann. Phys.} {\bf 243} 1 \\
  Pluha\v{r} Z, Weidenm\"uller H A, Zuk J A and Lewenkopf C H 1994
     {\it Phys. Rev. Lett.} {\bf 73} 2115           

\bibitem{Kadanoff84}
  Kadanoff L P and Tang C 1984 
     {\it Proc. Natl. Acad. Sci. USA} {\bf 81} 1276  \\
  Hannay J H and Ozorio de Almeida A M 1984 
     {\it J. Phys. A: Math. Gen.}   {\bf 17} 3429   
       
\bibitem{Ozorio98}
  Ozorio de Almeida A M, Lewenkopf C H and Mucciolo E R 1998
     {\it Phys. Rev. E} {\bf 58} 5693


\end{thebibliography}
\end{document}